\documentclass[english,aps,pra,reprint,showpacs]{revtex4-1}   
\usepackage[T1]{fontenc}	
\usepackage[latin9]{inputenc}	
\usepackage{geometry}		
\usepackage{graphicx}
\usepackage[above,below]{placeins}	
\usepackage{times}
\usepackage{amsmath}
\usepackage{multirow}

\usepackage[textwidth=4.5cm]{todonotes}

\begin{document}

\title{Active learning in pre-class assignments: Exploring the use of interactive simulations to enhance reading assignments}
\author{Jared B. Stang$^1$, Megan Barker$^2$, Sarah Perez$^3$, Joss Ives$^1$, and Ido Roll$^3$}
\affiliation{$^1$Dept. of Physics \& Astronomy, University of British Columbia, 6224 Agricultural Road, Vancouver, BC, V6T 1Z1\\
$^2$Dept. of Biological Sciences, Simon Fraser University, 8888 University Drive, Burnaby, BC, V5A 1S6\\
$^3$Centre for Teaching, Learning \& Technology, University of British Columbia, 214 - 1961 East Mall, Vancouver, BC, V6T 1Z1
}


\begin{abstract}
Pre-class reading assignments help prepare students for active classes by providing a first exposure to the terms and concepts to be used during class. We investigate if the use of inquiry-oriented PhET-based activities in conjunction with pre-class reading assignments can improve both the preparation of students for in-class learning and student attitudes towards and engagement with pre-class assignments. Over three course modules covering different topics, students were assigned randomly to complete either a textbook-only pre-class assignment or both a textbook pre-class assignment and a PhET-based activity. The assignments helped prepare students for class, as measured by performance on the pre-class quiz relative to a beginning-of-semester pre-test, but no evidence for increased learning due the PhET activity was observed. Students rated the assignments which included PhET as more enjoyable and, for the topic latest in the semester, reported engaging more with the assignments when PhET was included.
\end{abstract}

\maketitle
\section{Introduction}
Pre-class preparation positions students to get the most out of face-to-face class time \cite{Schwartz1998}. This is particularly important for active learning classrooms, where students rely on their prior learning to participate in peer discussion and construct their own further understanding of key concepts.   The standard approach to foster student preparedness is the assignment of pre-class reading. This is an evidence-based practice: in courses using pre-assignments coupled with quizzes, students complete the reading more consistently \cite{HeinerAJP2014,MarcellIJSoTL2008}, ask more---and more challenging---questions during class \cite{MarcellIJSoTL2008,Narloch2006}, and perform better on course assessments and exams \cite{HeinerAJP2014,Narloch2006,Dobson2008,Johnson2009}. Notably, a common thread among most studies is in the use of a relatively passive modality in pre-class assignments---textbook reading or watching videos.  While it is possible that students may actively engage---by taking notes, developing questions, and making connections with the material---there is no guarantee that they are using these types of strategies.  Additionally, they may have low intrinsic motivation to deeply engage, given the minimal opportunity for inquiry and exploration. As a possible solution, multimedia learning modules used for pre-lecture assignments were shown to promote learning and to improve student attitudes \cite{StelzerAJP2009,StelzerAJP2010}. Thus, there is a clear opportunity to deliberately use non-passive tools in the context of pre-class preparation assignments to support active engagement, inquiry, and other positive student outcomes.
%
%
%
%
%
%

Simulations offer one such tool.  The PhET Interactive Simulations have been carefully designed \cite{PerkinsPhysTeach2006} with dual goals of student engagement and learning. The results have been positive: PhETs promote learning, especially when students are exploring in a manner driven by their own questioning \cite{WiemanScience,AdamsJILC2008}. Additionally, the nature of PhET-based assignments that promote productive engagement has been investigated. In interviews, maximal engagement was seen in the presence of minimal---but non-zero---guidance \cite{AdamsAIP2008}. A separate in-class study provided students with either light-, moderate-, or heavy-guidance PhET-based assignments; students in the light guidance condition explored the PhET more and paid more attention to their interactions with the simulation \cite{ChamberlainChem2014}. Adams {\it et al.}~found that students can learn from PhETs at home, in an unstructured environment, and that moderate, question-driven guidance may be best in this context \cite{AdamsPERC2015}. 
%
%
%
%

\begin{table*}[htbp]
\caption{Pre- and post-test scores by topic. The pre-test was administered at the beginning of the semester, and the identical post-test occurred immediately after the pre-class assignment. $N$ is the number of students who completed all of the pre-test, post-test, and in-class test for the topic. A Wilcoxon signed-rank test shows that, for all topics, students learned during the pre-class assignments.}\label{tab1}
\begin{ruledtabular}
\begin{tabular}{lccccccccc}
 && \multicolumn{3}{c}{{\bf Pre-test}} & \multicolumn{3}{c}{{\bf Post-test}}  \\
{\bf Topic} & $N$ & {\bf Mean} & {\bf Median} & {\bf SD} & {\bf Mean} & {\bf Median} & {\bf SD} & $p$ & {\bf Effect size}\\
 \hline
Blackbody Radiation & 395 & 42\% & 33\% & 29\% & 53\% & 66\% & 29\% & $<.001$ & -.22  \\
Masses and Springs & 324 & 21\% & 0\% & 32\% & 43\% & 50\% & 40\% & $<.001$ & -.33 \\
Resonance & 342 & 31\% & 33\% & 28\% & 44\% & 33\% & 32\% & $<.001$ & -.22 
\end{tabular}
\end{ruledtabular}
\end{table*}

Our goal in this work is to leverage the capacity of PhET simulations to create productive engagement and learning to help better prepare students for class. We add active learning, rather than just passive reading, to the pre-class assignments by including inquiry-oriented PhET-based activities, giving students an opportunity to explore within the constrained environment of the PhET. It is plausible that this type of hands-on exploration can promote a conceptual understanding that helps students piece together the knowledge they encounter in the textbook reading and during in-class instruction. Specifically, we investigate if the use of PhET-based activities in conjunction with pre-class reading assignments can improve both the preparation of students for in-class learning and student attitudes towards and engagement with pre-class assignments.

\section{Method}

This work took place at a large research-intensive university in the Pacific Northwest.  The study was conducted in a one-semester, multi-section first-year physics course for engineering students on the subject of thermodynamics and periodic motion. Each of three lecture sections for the course contained about 260 students, with a total of 779 students in the course.  Within this course, for each of three different modules (Blackbody Radiation, Masses and Springs, and Resonance), we created assignments with two pieces, based on either the textbook or a PhET simulation on the same topic.  The text pre-reading was a standard pre-class reading assignment, which included targeted questions that directed students to engage with specific passages, vocabulary, concepts, and figures in the book \cite{HeinerAJP2014}.  The PhET assignment directed students to interact with the relevant PhET simulation by providing a hyperlink to the PhET and prompting them to explore particular relationships within the simulation and record their observations in an open-response text box. For example, for the topic Masses and Springs, the prompt was: ``Explore the relationships between the spring softness, mass, amplitude, and period of oscillation. In the space below, describe 2-3 interesting things you noticed." In comparison to the different levels of guidance described in \cite{ChamberlainChem2014}, our PhET assignments were closest to their ``light guidance'' condition. While textbook-based pre-reading assignments are standard in this course, and PhETs have been recommended to students in previous iterations of the course, the explicit use of the PhET assignment was novel in this course.

These two styles of activities---Text and PhET---engage the students differently with a topic: reading a textbook transmits knowledge whereas PhET simulations reveal knowledge through exploration of the concepts. We sought to test the hypotheses that 1) inclusion of the PhET assignment would improve student performance and engagement over a text-only pre-reading, and 2) when both PhET and text are assigned, the relative order of the two would impact student performance. Students were randomly assigned to three treatment groups: 
 text pre-reading assignment alone; PhET activity before text pre-reading; and text pre-reading before PhET activity.  The experiment was repeated over three course modules and each student group experienced each treatment once over the course of the study.  To ensure the correct order of completion, students were not given access to their second part of the assignment (text or PhET) until they had completed the first.

Student performance and perspectives data were collected from these students throughout the semester.  To assess baseline student understanding ("pre-test"), students completed an online 8-question multiple choice pre-test early in the semester, prior to any exposure to the study topics. The test, graded for participation, included three questions each related to Blackbody Radiation and Resonance and two questions related to Masses and Springs.  Subsequently, within each module, student learning was assessed immediately following the experimental treatment, but before in-class instruction ("post-test"), and again following in-class instruction ("in-class test").  The post-test was completed as part of a normal online pre-class quiz, and consisted of the identical questions from the cognate topics in the pre-test.  The in-class test was a series of 3-4 clicker questions, given in class 2-3 days later.  Students were directed to work independently for these questions.  Student perspectives data were collected using an online survey (for participation grades) administered with the post-test.  Students were asked 1) to self-report the amount of time  spent on both parts of the assignment, and 2) their perceived enjoyment and learning from the entire pre-class assignment. The perspectives data were collected only for the latter two modules (Masses and Springs; Resonance).

For the analysis of learning outcomes, only students who completed all performance assessments for a topic (pre-test, post-test, and in-class test) were included in the study cohort. For each student, a raw score (percent correct) was computed for each assessment, for each topic. Only students who completed all survey questions for a topic were included in the analysis of perceptions ($N=408$ for Masses and Springs and $N=407$ for Resonance).
\section{Results}
\subsection{Learning}
%
Over all three topics, students learned during the pre-class assignment, as measured by an increase in score from pre-test to post-test. These results are shown in Table~\ref{tab1}. 

To evaluate whether or not student learning during the pre-class assignment depended on the treatment (text reading alone, PhET activity then text reading, or text reading then PhET activity), the following linear model was used,
\begin{align}\label{model}
&\textrm{Post-test}_{ijk}=\beta_0+\beta_{1}\times\textrm{Pre-test}_{ij}+\beta_{2,j}\times\textrm{Topic}_j\\
&\quad\;\;+\beta_{3,k}\times\textrm{Treatment}_k+\beta_{4,jk}\times\textrm{Topic}_j\textrm{Treatment}_k+\varepsilon_i,\nonumber
\end{align}
where $\textrm{Post-test}_{ijk}$ is the post-test score of student $i$ on $\textrm{Topic}_j$, $\textrm{Pre-test}_{ij}$ is the pre-test score for student $i$ on $\textrm{Topic}_j$, $\textrm{Topic}_j$ is a categorical variable representing the topic (Blackbody Radiation, Masses and Springs, or Resonance), $\textrm{Treatment}_k$ is a categorical variable representing the treatment condition, and $\varepsilon_i$ is a random intercept for student $i$ which accounts for differences in students. We include a $\textrm{Topic}_j\textrm{Treatment}_k$ interaction term to take into account that the different PhETs may promote learning differently.

There was no significant effect for treatment, when controlling for pre-test and topic, $F(2, 695) = 1.1$, $p=.33$, nor for the interaction between treatment and topic, $F(2, 695) = 0.041$, $p=.96$.  As expected, pre-test was a highly significant predictor of learning, $F(1, 695) = 50$, $p<.001$, $\eta^2 = .067$, which corresponds to a medium effect size. 
 
To evaluate whether or not the type of pre-class assignment influenced student learning in-class, we used a similar model as (\ref{model}), replacing post-test with the in-class clicker test as the dependent variable. As for the post-test, no evidence for an effect of the treatment on the in-class test results was found.
%
%
%
\subsection{Engagement and perceptions}
Over all treatment conditions, the students reported spending a mean time of 34.4 min (median = 30 min, SD = 29.0 min) on the textbook reading. Over the two conditions involving PhET assignments, students reported spending a mean time of 10.3 min (median = 10 min, SD = 10.7 min) on the PhET activity. For the topic Resonance, when their pre-class assignment included the PhET activity, students reported spending more time overall on the pre-class assignment (Table~\ref{tab2}).
\begin{table}[htbp]
\caption{Time on task by topic and condition. The overall time for each PhET treatment is compared to the text-only treatment with a Mann-Whitney test, with effect size (rank-biserial correlation) $r$. For the Resonance topic, when the PhET activity was included, students reported longer time-on-task with the pre-class activity.\label{tab2}}
\begin{ruledtabular}
\begin{tabular}{llcccccc}
 & \multicolumn{5}{c}{{\bf Overall Reported Time on task (min)}} \\
{\bf Treatment} & $N$ & {\bf Mean} & {\bf SD} & $p$ & $r$\\
 \hline
\textbf{Masses \& Springs} \\
Text only & 144 & 45.9 & 31.7 \\
PhET then text & 137 & 51.4 & 42.6 & .47 & -.050 \\
Text then PhET & 127 & 44.0 & 29.6 & .65 & .032\\
\hline
\textbf{Resonance} \\
Text only & 128 & 29.5 & 19.8 \\
PhET then text & 143 & 35.5 & 24.3 & .009 & -.18\\
Text then PhET & 136 & 40.8 & 33.8 & <.001 & -.29
\end{tabular}
\end{ruledtabular}
\end{table}

The time students spent engaged with the PhET activity did not seem to depend on the order of the assignments: Students did spend slightly more time on PhET if it was before the Text assignment, though not significantly, Mann-Whitney $p=.087$, $r=.08$. To investigate if engagement with the PhET assignment promoted learning during the pre-class assignment, the model in (\ref{model}) was run with PhET engagement as a predictor. No evidence for an effect of PhET engagement on learning from the pre-class assignment was found (as measured on either the post-test or in-class test).

Students reported that they enjoyed the pre-class assignments and found them beneficial to their learning. Splitting by treatment, 71\% of students whose assignment included a PhET activity rated the pre-class assignment as enjoyable compared to 62\% of students in the textbook-only group. Overall, 74\% of students found the pre-class assignments beneficial to their learning, with no difference in perceived learning benefit across treatment groups.

Within the two PhET treatments, there is a correlation between engagement with the PhET activity and how beneficial to their learning and enjoyable students found the pre-class assignment (Table~\ref{tab3}): Students who reported the assignment as beneficial to their learning (enjoyable) spent more time with the PhET activity than students who reported the assignment as not beneficial to their learning (not enjoyable).

\begin{table}[htbp]
\caption{Time on task with the PhET activity split on perceptions of the pre-class assignment. By a Mann-Whitney test, students who reported the assignment as beneficial to their learning (enjoyable) spent more time with the PhET activity than students who reported the assignment as not beneficial to their learning (not enjoyable).\label{tab3}}
\begin{ruledtabular}
\begin{tabular}{lccccc}
 & \multicolumn{5}{c}{{\bf PhET Reported Time on task (min)}} \\
{\bf Perception} & $N$ & {\bf Mean} & {\bf SD} & $p$ & $r$\\
 \hline
Beneficial to learning & 404 & 11.1 & 11.8\\
Not beneficial to learning & 139 & 7.96 & 6.25 & <.001 & -.24\\\hline
Enjoyable & 384 & 10.7 & 8.35\\
Not enjoyable & 159 & 9.20 & 14.9 & <.001 & -.22
\end{tabular}
\end{ruledtabular}
\end{table}

\section{Discussion}
From pre- to post-test, students learned during the pre-class assignments, with a pedagogically significant effect size. This indicates that the assignments were indeed successful in helping prepare students for class. Although we hypothesized that PhETs could contribute to learning during the pre-class assignments, our analysis revealed no effect on our measures of learning relative to the textbook-only assignments. It could be that, for the at-home pre-class assignment, our light guidance did not provide the scaffolding students would need to productively explore within the PhET activity. If this were the case, the Text then PhET treatment, for which reading the textbook before the PhET activity may provide extra scaffolding, may have had a better chance of providing the structure students would need for productive engagement with the PhET. However, no learning effect was seen whether students were constrained to do the PhET activity before or after the textbook reading, so it may be that, no matter the order, the level of guidance was too light for this assignment. In their study of PhET activities in unstructured environments, Adams {\it et al.}~were unable to create effective activities with light guidance \cite{AdamsPERC2015}; it may be that more direct instruction is necessary for pre-class activities as well.

It is possible that, although the assignment was constrained, students referred to the textbook through all parts of the assignment. Then, by also having a textbook assignment, productive exploration may have been undermined. Because the textbook part of the assignment is important to students' preparation for the class---so that they have encountered basic definitions and concepts---it may be difficult to disentangle the learning specific to the PhET activity. 
 It may also be that our assessment questions, while appropriate for the topic in general, did not target the specific concepts addressed in the PhETs, or that these particular PhETs did not provide exploration opportunities at a level appropriate for this audience.

Including a PhET activity in the pre-class assignment influenced both student engagement with and perceptions of the learning benefits of the assignments. When PhET was included, more students reported the pre-class assignments as enjoyable and as many students as the text-only condition rated them as beneficial to their learning. Additionally, when PhET was included in the Resonance topic, students reported spending 5--10 minutes longer with the assignments. This is notable, as Resonance was the most difficult of the topics and was encountered late in the semester, when competing priorities and general fatigue may have prevented students from engaging. The affective benefits of including PhET may have helped to sustain the motivation of these students.

Interestingly, the reported time spent on the PhET activity did not seem to depend on the ordering of activities. If completing the textbook reading before the PhET activity affected how students interacted with the PhET, it did not show up here. Overall, students in both PhET conditions reported spending marginally less time on the textbook part of the assignment. This is odd, because students in both the Text or Text then PhET treatment did not know if they were going to have a PhET activity, so we would expect the textbook engagement for these treatments to match. Since their perceived time spent on the assignments was recorded after finishing all activities, it appears that having PhET as part of the assignment altered how students thought about the assignment as a whole.

There was an interesting correlation between time spent on the PhET and perceptions of the pre-class assignments. It could be that students who enjoyed the assignment, topic, or PhET more were pre-disposed to spend more time with the simulation, or it could be that students who spent more time with the PhET ended up enjoying the assignment more. More study is needed to understand how student engagement and attitudes interact. No effect of PhET engagement on learning was observed, indicating that the extra time spent did not translate into learning gains on our measures.

\section{Conclusion}
PhETs are not a magic bullet: They have been shown to promote learning with carefully designed activities in particular contexts. This report shows that even purposeful inclusion of PhETs in pre-class assignments does not automatically translate into learning. In the case of pre-class assignments, more study is required to see if it is possible to design supplemental PhET activities that improve student preparation beyond that from a text-only assignment. The inclusion of PhET did have an effect on both student engagement with the pre-class assignment and student perceptions of the assignments. These affective results suggest that PhETs did contribute meaningfully to these assignments. Sustaining student motivation can be a challege, so the potential for PhETs to contribute in this domain is an important consideration.

\acknowledgments{We thank Don Witt and Andrzej Kotlicki for early conversations and support in this project.}



\begin{thebibliography}{99}  

\bibitem{Schwartz1998} D.L. Schwartz and J.D. Bransford, Cogn. Instr. {\bf 16}, 475 (1998).

\bibitem{HeinerAJP2014} C.E. Heiner {\it et al.}, Am.~J.~Phys.~{\bf 82}, 989 (2014).

\bibitem{MarcellIJSoTL2008} M. Marcell, International Journal for the Scholarship of Teaching and Learning {\bf 2}, 7 (2008).

\bibitem{Narloch2006} R. Narloch {\it et al.}, Teach.~Psychol. {\bf 33}, 109 (2006).

\bibitem{Dobson2008} J.L. Dobson, Adv.~Physiol.~Educ. {\bf 32}, 297 (2008).

\bibitem{Johnson2009} B.C. Johnson and M.T. Kiviniemi, Teach.~Psychol. {\bf 36}, 33 (2009). 

\bibitem{StelzerAJP2009} T. Stelzer {\it et al.}, Am.~J.~Phys.~{\bf 77}, 184 (2009).

\bibitem{StelzerAJP2010} T. Stelzer {\it et al.}, Am.~J.~Phys.~{\bf 78}, 755 (2010).

\bibitem{PerkinsPhysTeach2006} K. Perkins {\it et al.}, Phys.~Teach.~{\bf 44}, 18 (2006).

\bibitem{WiemanScience} C.E. Wieman {\it et al.}, Science {\bf 322}, 682 (2008).

\bibitem{AdamsJILC2008} W.K. Adams {\it et al.}, Journal of Interactive Learning Research {\bf 19(3)}, 397 (2008).

\bibitem{AdamsAIP2008} W.K. Adams {\it et al.}, AIP Conf.~Proc.~{\bf 1064}, 1 (2008).

\bibitem{ChamberlainChem2014} J.M. Chamberlain {\it et al.}, Chem.~Educ.~Res.~\& Practice {\bf 15}, 628 (2014).

\bibitem{AdamsPERC2015} 
W.K. Adams {\it et al.}, \textit{2015 PERC Proceedings} (2015). DOI: 10.1119/perc.2015.pr.001




\end{thebibliography}
\end{document}